\documentstyle[12pt, cite, epsfig, a4]{article}

\newcommand{\la}{\langle}
\newcommand{\ra}{\rangle}

\newcommand{\beq}{\begin{equation}}
\newcommand{\eeq}{\end{equation}  }

\def\E{\mbox{e}^+\mbox{e}^-}

\begin{document}

\begin{center}

{\Large\bf Bunching Parameter and Intermittency
 in High-Energy Collisions}

\vspace{2.0cm}

{\large S.V.Chekanov  and
V.I.Kuvshinov}

\bigskip

{\it Institute of Physics, Academy of Sciences of Belarus\\
F.Scaryna Av. 70, 220072 Minsk, Belarus}

\vspace{1cm}

{\large\em Published in Acta Phys. Pol. B25 (1994) 1189}

\end{center}

\vspace{1cm}

\begin{abstract}
We introduce the parameter of bunching for
an analysis of the intermittent structure of
multihadron production in high-energy collisions following an
analogy with photon
counting in quantum optics. A power-law
singularity is shown to exist for second-order
bunching parameters
in small phase-space intervals for the case of
a monofractal structure of the multiplicity
distribution and a similar form for the
high-order parameters for the case of multifractality.
The approximation of the high-order bunching parameters by the second-order
provides a good description of the anomalous fractal dimensions for
a number of experimental data with multifractal behavior.
\end{abstract}

\section{Introduction}
\label{sec:bp1int}

The idea of applying stochastic methods developed to study photon counting
statistics in quantum optics to particle production processes has been used
for a long time \cite{1,2}. 
At present, a systematic and
careful investigation of multihadron production by the application of
methods borrowed from 
quantum optics is very useful because there is a large analogy
between these fields of physics. For instance, the interpretation
of multiplicity distributions in terms of hadronic field states by
analogy to photon counting \cite{3,4} and the study of squeezed gluon
states \cite{5} seem to be  important
directions for theoretical research. The problem of damping of the
statistical noise and the concept of 
factorial moment analysis to study multihadron
production \cite{6} have long been known in quantum optics \cite{7}. 
Correlators in terms
of moments \cite{8} have an analogous form in quantum optics as well \cite{9}.

The purpose of the present paper is to extend some methods of continuous
quantum
measurement in quantum optics to high-energy physics. 
We introduce the bunching
parameter for the analysis of fractal structure 
of multihadron production.
For intermittent structure of the multiplicity distribution, 
a non-trivial behavior is obtained  for small phase-space intervals.

\section{The bunching parameter}
\label{sec:bp11}

In the theory of continuous quantum measurement, 
the bunching parameter $\check\eta_{q}(\delta t)$ of order $q$ 
for an one-mode photon field
can be expressed  in  terms of
the probability $P_{n}(\delta t)$ to have $n$ photons in the time interval
$\delta t$ \cite{10} in the form
\beq
\check\eta_{q}(\delta t)=\frac{q}{q-1}\frac{P_{q}(\delta t)P_{q-2}(\delta t)}
{\left(P_{q-1}(\delta t)\right)^{2}}, \qquad q>1.
\label{1z}
\eeq
This parameter determines how the probability to detect $q$
photons in $\delta t$ changes
relatively to the probability to detect $q-1$ 
photons in the same time interval.
If the source of light is completely coherent,
then $\check\eta_{q}(\delta t)=1$.
The corresponding multiplicity distribution is the Poisson one.
A radiation field is said to be statistically antibunched in order $q$ if
$\check\eta_{q}(\delta t)<1$. When $\check\eta_{q}(\delta t)>1$, then
it is said to be bunched in $\delta t$. For a wide class of states,
the bunching parameters are independent of the time interval [10].

By analogy with (\ref{1z}), let us consider the bunching parameters (BPs)
$\eta_{q}(\delta )$
for the multiplicity distributions of 
secondary particles produced in high-energy
interactions
\beq
\eta_{q}(\delta )=\frac{q}{q-1}\frac{P_{q}(\delta )P_{q-2}(\delta )}
{\left(P_{q-1}(\delta )\right)^{2}}, \qquad q>1,
\label{2z}
\eeq
where $P_{n}(\delta )$ is the probability to have $n$ particles in the 
phase-space interval
$\delta$ defined in rapidity, azimuthal angle,
transverse momentum or a
(multi-dimensional) combination of these variables.

There is a large class of distributions which 
has  $\delta$-independent BPs.
By applying formula (\ref{2z}),
any multiplicity distribution can be expressed as
\beq
P_{n}(\delta )=
P_{0}(\delta )\frac{\lambda^{n}(\delta )}{n!}\prod_{m=2}^{n}\left[
\eta_{m}(\delta )\right]^{n+1-m}, \qquad n>1,
\label{3z}
\eeq
where $\lambda (\delta )=P_{1}(\delta )/P_{0}(\delta )$. If $\eta_{q}
(\delta )$ is independent of $\delta$, 
one gets the following  general form of the generating
function for a such distribution
\beq
G(z, \delta )\equiv\sum_{n=0}^{\infty}z^{n}P_{n}(\delta )=
G(z=0, \delta )Q\left(z\lambda (\delta )
\right),
\label{4z}
\eeq
where $G(z=0, \delta )=P_{0}(\delta )$, the $Q(z\lambda (\delta ))$
is some analytic
function of the auxiliary variable $z$ multiplied by a function
$\lambda (\delta )$ under the 
condition $Q(\lambda (\delta))=1/G(z=0, \delta )$.
It is easy to see that condition (\ref{4z}) is fulfilled for
well-known distributions, such as the Poissonian, binomial  and
geometric ones. The case of a negative-binomial 
distribution will be discussed
later.

Let us remind that the observed behavior
of the normalized factorial moments (NFMs)
\beq
F_{q}(\delta )\equiv\frac{\langle n^{[q]}\rangle }{\langle n\rangle ^{q}}
\propto{\delta}^{-d_{q}(q-1)},
\qquad \delta \to 0 , \qquad q\ge 2
\label{5z}
\eeq
is a straightforward manifestation of  nonstatistical intermittent
fluctuations
in the distribution of secondary particles produced in high-energy
interactions \cite{6,11,12}. 
In (\ref{5z}) $n$ denotes the number of particles in $\delta$,
$n^{[q]}\equiv n(n-1)\dots (n-q+1)$, $\langle\ldots \rangle$ is the 
average over all events.
The right side of (\ref{5z}) represents the definition of the intermittent
behavior characterized by an anomalous fractal dimensions (AFDs) $d_{q}$
depending on the rank $q$ of the 
NFMs  for multifractal behavior and $d_{q}=const$
for monofractality.

Now we shall prove the following statement: an inverse power 
$\delta$-dependence of the
second-order BP  
and $\delta$-independence of
high-order BPs are necessary and sufficient conditions for the monofractal
behavior of AFDs. An inverse power $\delta$-dependence of all
BPs is necessary and sufficient for  multifractality.

{\it Sketch of a proof}. 
By applying (\ref{3z}), for the NFMs in terms of BPs,
we have
\beq
F_q(\delta )=\frac{P_0(\delta )}{\la n\ra ^q}\sum_{n=q}^{\infty}
\frac{n^{[q]}}{n!}\lambda^{n}(\delta )
\prod_{m=2}^{n}\left[\eta_{m}(\delta )\right]^{n+1-m},
\label{6z}
\eeq
\beq
\la n\ra =P_{0}(\delta )\lambda (\delta )+P_{0}(\delta )\sum_{n=2}^{\infty}
\frac{\lambda^{n}(\delta )}{(n-1)!}\prod_{m=2}^{n}\left[\eta_{m}(\delta )
\right]^{n+1-m}.
\label{7z}
\eeq
Assuming the approximate proportionality
of $\la n\ra$ and $\delta$ at small $\delta $ and  the condition
$P_{0}(\delta )\to 1$, for $\delta\to 0$, we have the following
approximate expression for the small interval $\delta $
\beq
F_{q}(\delta)\simeq\prod_{m=2}^{q}\left[\eta_{m}(\delta )\right]^{q+1-m}.
\label{8z}
\eeq
In the case of a power-law dependence of NFMs (\ref{5z}) with  monofractal
behavior of AFDs $d_{q}=d_{2}=const$, 
we must require the following properties of BPs 
\beq
\eta_{2}(\delta )\propto\delta^{-d_{2}},\qquad 0<d_{2}<1,
\label{9z}
\eeq
\beq
\eta_{s}(\delta )\simeq const, \qquad s>2.
\label{10z}
\eeq
From expression (\ref{8z}), one can conclude 
that for multifractality a
power-law singularity of the BPs of the order $q=s>2$ is necessary.
Using (\ref{8z}), it is easy to show the sufficient conditions of an
inverse power-law behavior
of BPs for both the monofractal and multifractal cases.

Now we have the possibility to write down the general form of
the generating function $G^{\mathrm{mon}}(z, \delta )$
for the multiplicity distribution
with monofractal behavior of AFDs
\beq
G^{\mathrm{mon}}(z, \delta )=G^{\mathrm{mon}}(z=0, \delta )
\left(1+\eta_{2}^{-1}(\delta )Q^{\mathrm{mon}}\left(
\lambda (\delta )\eta_{2}(\delta )z\right)\right),
\label{11z}
\eeq
where $\eta_{2}(\delta )$ is defined by (\ref{9z}), $Q^{\mathrm{mon}}$
is some analytic
function with variable $\lambda (\delta )\eta_{2}(\delta )z$
with the following conditions
\beq
Q^{\mathrm{mon}}\left(\lambda (\delta )\eta_{2}(\delta )z=0\right)=0,
\label{12z}
\eeq
\beq
Q^{\mathrm{mon}}\left(\lambda (\delta )\eta_{2}(\delta )z=\lambda (\delta )
\eta_{2}(\delta )\right)=\eta_{2}(\delta )\left(
\frac{1}{G^{\mathrm{mon}}(z=0)}-1\right).
\label{13z}
\eeq
The general formal form of the generating function for multifractal
behavior one can obtain from (\ref{3z}).

\section {The BPs for the negative-binomial distribution}
\label{sec:bp12}

Since a few years, many high-energy multiparticle data at various energies
have been successfully fitted by the negative-binomial distribution (NBD)
\cite{13,14,15,16} with the generating function
\beq
G^{\mathrm{NBD}}(\delta y, z)
=\left(1+\frac{\la n(\delta y)\ra }{k(\delta y)}(1-z)\right)^
{-k(\delta y)},
\label{14z}
\eeq
where $\la n(\delta y)\ra$ is the average multiplicity of final hadrons in
the restricted rapidity (or pseudorapidity) 
interval $\delta y$ and $k(\delta y)$
is a positive parameter. If $k(\delta y)$ does not depend on $\delta y$,
we do not have  any fractal type of behavior for 
the NFMs of this distribution.
Indeed, in this case one
can rewrite the generating function in the form (\ref{4z}).

In the general case, the BPs of NBD are given by the expression
\beq
\eta_{q}^{\mathrm{NBD}}(\delta y)=\frac{k(\delta y)+q-1}{k(\delta y)+q-2},
\qquad q=2,\ldots \infty .
\label{15z}
\eeq
Let us assume that $k(\delta y)\propto \delta y^{d_{2}}$.
In this case,
$\eta_{2}^{\mathrm{NBD}}(\delta y)\propto\delta y^{-d_{2}}$ and
$\eta_{s}^{\mathrm{NBD}}(\delta y)\simeq const$, 
for $s>2$ at small $\delta y$.
According to Sect.~\ref{sec:bp11}, one gets 
the monofractal type of behavior
for AFDs i.e. $d_{q}=d_{2}=const.$.
Such a monofractal behavior has already been discussed in \cite{17,18}.
This analysis only illustrates the simplicity of 
our approach to intermittency
in terms of BPs.

\section{The L\'{e}vy-law approximation}
\label{sec:bp13}

In this 
section we shall show the possible behavior of the bunching parameters
in rapidity bins for different high-energy collisions.

At the beginning, let us note that for an investigation of intermittency in
rapidity bins $\delta y$ one usually averages \cite{6} the factorial moments
over all bins of equal width $\delta y$ normalizing to the overall average
number per bin ${\overline {\la n\ra }}\equiv\sum_{m=1}^{M}\la n_{m}\ra /M$, 
where $\la n_{m}\ra $ is the average
multiplicity in the $m$th bin, $M=Y/\delta y$, $Y$ being the
full rapidity interval,
\beq
\bar F_{q}(\delta y)\equiv \frac{1}{M}\sum_{m=1}^{M}
\frac{\la n_{m}^{[q]}\ra }{{\overline {\la n\ra }}^{q}}
\simeq C_{q}{(\delta y)}^{-d_{q}(q-1)},
\label{16z}
\eeq
where $C_{q}$ is some constant.
Similarly, one can introduce the BPs by averaging the
probability $P_{n}^{m}(\delta y)$ for the $m$th bin over all $M$ bins,
\beq
\bar\eta_{q}(\delta y)=
\frac{q}{q-1}\frac{\bar P_{q}(\delta y)\bar P_{q-2}(\delta y)}
{\left(\bar P_{q-1}(\delta y)\right)^{2}}, \qquad q>1,
\label{17z}
\eeq
where $\bar P_{n}(\delta y)\equiv\frac{1}{M}\sum_{m=1}^{M}P_{n}^{m}(\delta y)$.

Following the same procedure as 
in Sect.~\ref{sec:bp11}, we can see that the approximate
expression for the NFMs (\ref{16z}) in  terms of the 
BPs (\ref{17z}) has  the same form for
$\delta y\to 0$ as (\ref{8z}), if we substitute $\bar F_{q}(\delta y)$
for $F_{q}(\delta )$ and $\bar\eta_{s}(\delta y)$ for
$\eta_{s}(\delta )$. Then, we have
\beq
\bar\eta_{2}(\delta y)\simeq\bar F_{2}(\delta y),\qquad\bar\eta_{s}(\delta y)
\simeq\frac{\bar F_{s}(\delta y)\bar F_{s-2}(\delta y)}
{\left(\bar F_{s-1}(\delta y)\right)^2},
\label{18z}
\eeq
where $\bar F_{1}(\delta y)=1$ and  $s>2$. 
Using (\ref{16z}) and (\ref{18z}), we obtain the following
expression for the BPs
\beq
\bar\eta_{2}(\delta y)\simeq C_{2}\delta y^{-\beta_{2}},\qquad
\bar\eta_{s}(\delta y)
\simeq\frac{C_{s}C_{s-2}}{\left(C_{s-1}\right)^2}\delta y^
{-\beta_{s}},
\label{19z}
\eeq
where $C_{1}=1$ and
\beq
\beta_{2}=d_{2}
\label{20z}
\eeq
\beq
\beta_{s}=d_{s}(s-1)+d_{s-2}(s-3)-2d_{s-1}(s-2),\qquad s>2.
\label{21z}
\eeq
Note that we can obtain  (\ref{21z}) using the approximation
$\la n_{m}^{[q]}\ra\simeq q!P_{q}^{m}(\delta y)$ 
for ${\overline {\la n\ra }}\ll 1$
(a similar analysis of factorial moments in terms of the probabilities for
one bin can be found in [19]).

For an analysis of the parameters $\beta_{n}$,
we shall use the
L\'{e}vy-law approximation for the 
AFDs \label{20} which has been introduced to describe
random cascading models,
\beq
d_{q}=\frac{d_{2}}{2^{\mu}-2}\frac{q^{\mu}-q}{q-1},
\label{22z}
\eeq
with the L\'{e}vy index $\mu$. The L\'{e}vy index is known as a 
degree of multifractality
and allows a natural interpolation between the monofractal case $(\mu =0)$
and multifractality $(\mu >0)$. The case $\mu =2$ $(d_{q}=qd_{2}/2)$
corresponds to the log-normal approximation.
Substituting (\ref{22z}) into (\ref{21z}), one gets the following expression
\beq
\beta_{q}=d_{2}E_{q},
\qquad E_{q}\equiv\frac{q^{\mu}+{(q-2)}^{\mu}-2{(q-1)}^{\mu}}
{2^{\mu}-2}.
\label{23z}
\eeq
In the limit of monofractal behavior of the AFDs ($\mu =0$), we have
$\beta_{s}=0$ for $s>2$, i.e. the high-order BPs are independent of
$\delta y$. Then, the values of $\bar\eta_{s}(\delta y)$
are completely determined by the coefficients $C_{q}$. Note, that in the
case of multifractality, the values of $E_{q}$ are
positive for all $q$ and the BPs
increase indefinitely for $\delta y\to 0$. Thus, in the case of multifractal
behavior, one can speak  of  a strong  bunching of particles in $\delta y$.
For the log-normal approximation ($\mu =2$), we obtain $E_{q}=1$,
$\beta_{q}=d_{2}$, i.e all 
bunching parameters have the same power-law behavior
$\bar\eta_{q}\propto\delta y^{-d_{2}}$.

Thus, there are two important limiting  cases 
which correspond to  monofractality
and log-normal approximation for multifractality :
\beq
\mu =0,\qquad\bar\eta_{2}(\delta y )\propto\delta y^{-d_{2}},
\qquad \bar\eta_{s}=const.,\qquad\hbox{for all}\qquad s\ge 3,
\label{24z}
\eeq
\beq
\mu =2,
\qquad\bar\eta_{q}(\delta y)\propto\delta y^{-d_{2}},\qquad\hbox{for all}
\qquad q\ge 2.
\label{25z}
\eeq
In a real physical situation,
the L\'{e}vy index $\mu$ is different
for different reactions \cite{12,20,21,22} and, strictly speaking,
it is not equal to an integer value:

(i) Nucleus-nucleus reaction $S-AgBr$: The L\'{e}vy index is $0\le\mu <0.55$
(in fact, it is almost a monofractal system) \cite{20}. 
The value of $E_{q}$ is almost zero
and the BPs approximately are $\delta y$-independent.
This behavior is typical for the 
intermittency at second-order phase transition and thus
has been advocated \cite{23} in favor of the formation of a quark-gluon plasma.

(ii) $\E$, pA, AA, hh reactions with $\mu\simeq 1.3-1.6$
\cite{22} corresponding to the parameter $0<E_{q}<0.7$.
In the case of $\mu$p deep inelastic scattering, 
the L\'{e}vy index is largest,
$\mu\simeq 2.6-3.7$ \cite{20,21} and $1<E_{q}<5$.
In these cases, we have a power-law singularity in the behavior
of the BPs.

\begin{figure}[hhh]
\begin{center}
\mbox{\epsfig{file=f1p.eps, width=13.0cm}}
\caption[f1p]
{\it
Experimental data for the behavior of AFDs [24-26].
Continuous lines show
fits using (\ref{29z}).}
\label{f1p}
\end{center}
\end{figure}

\section{Simple approximation of the high-order BPs}

Using
only one free parameter $\mu$, the L\'{e}vy law approximation  allows a simple
description of multifractal properties of random cascade models. 
However, using the interpretation of
intermittency via the BPs, we can make some approximation of high-order
BPs in order to obtain a more 
simple linear expression for the AFDs, maintaining
the number of free parameters.

Let us make the assumption that the high-order BPs can be
expressed in terms of the second-order BP and a constant $r>0$ as
\beq
\bar\eta_{s}(\delta y)=\left(\bar\eta_{2}(\delta y)\right)^{r},
\qquad s>2
\label{26z}
\eeq
with
\beq
\bar\eta_{2}(\delta y)\propto\delta y^{-d_{2}}.
\label{27z}
\eeq
For the given case,
the multiplicity distribution with multifractal behavior has the
following form  ($n>1$)
\beq
\bar P_{n}(\delta y)=\bar P_{0}(\delta y)\frac{\bar\lambda^{n}(\delta y)}{n!}
\left[\bar\eta_{2}(\delta y)\right]^{n-1+\frac{r}{2}(n-n^2+8)(1-\delta_{2n})},
\label{28z}
\eeq
where $\bar\lambda (\delta y)=\bar P_{1}(\delta y)/ \bar P_{0}(\delta y)$
and $\delta_{22}=1$, $\delta_{2n}=0$ for $n\ne 2$.
Using (\ref{18z}), (\ref{26z}-\ref{27z}), 
the AFDs of such distribution are given by the linear expression
\beq
d_{q}=d_{2}(1-r)+d_{2}r\frac{q}{2}.
\label{29z}
\eeq
This linear approximation, in our opinion, is 
very interesting, because
it allows interpolation between the monofractal case ($r=0$) and
the log-normal approximation ($r=1$) as does 
the L\'{e}vy-law approximation (\ref{22z}).
The results of fits of some experimental data [24-26] 
by the expression (\ref{29z}) are
presented in Figure ~\ref{f1p}. 
This analysis gives good agreement with the experimental
data. Thus, the  approximation (\ref{26z}) 
of high-order BPs by the second-order
is valid for such reactions.

\section{Conclusion}

We have introduced the bunching parameters for the analysis of multiparticle
production in high-energy physics by analogy with the theory of continuous
quantum measurement  for one-mode photon fields.
It is shown that an inverse power-law singularity of the second-order BP
leads to  monofractal behavior of the AFDs at  small rapidity intervals
if high-order BPs are independent of phase-space intervals.
Using a
such  dependence of the second-order BP on the phase-space interval size,
we have found a general form of the generating function with monofractality.
For multifractality,
an inverse power-law singularity for all order BPs is necessary and sufficient.
Using the experimental data, we can conclude that, for $\delta y\to 0$, 
the majority of reactions indeed  show  a strong bunching of particles in 
all orders.

We have shown that the investigated
experimental behavior of the AFDs can be understood
as a simple approximation of the high-order BPs in  terms of the second order.
We believe this to be an important conclusion as it leads to a description
of the multifractal multiplicity distribution with a minimum
number of free parameters.

The use of BPs is interesting because it gives a general answer
to the problem of finding a multiplicity 
distribution leading to intermittency.
This method is also interesting since it may provide 
a link between theory of
continuous quantum measurement and the investigations of multifractal
structure of multiplicity distributions in particle collisions at
high energies. It, furthermore, grants 
the possibility to analyze the intermittency
phenomenon in quantum optics.

\newpage
\def\refname{

\end{document}